\documentclass[aps,prc,twocolumn,floatfix,nofootinbib]{revtex4}
\usepackage{amsmath,graphicx}
\begin{document}


\title{New relativistic dissipative fluid dynamics from kinetic theory}

\author{Amaresh Jaiswal, Rajeev S. Bhalerao, and Subrata Pal}
\address{Tata Institute of Fundamental Research,
Homi Bhabha Road, Mumbai 400005, India}

\date{\today}

\begin{abstract}

Starting with the relativistic Boltzmann equation where the collision
term is generalized to include nonlocal effects via gradients of the phase-space
distribution function, and using Grad's
14-moment approximation for the distribution function,
we derive equations for the relativistic dissipative fluid
dynamics. We compare them with the corresponding equations obtained in
the standard Israel-Stewart and related approaches. Our method
generates all the second-order terms that are allowed by symmetry,
some of which have been missed by the traditional approaches
based on the 14-moment approximation, and the
coefficients of other terms are altered. The first-order or
Navier-Stokes equations too get modified. Significance of these
findings is demonstrated in the framework of one-dimensional
scaling expansion of the matter formed in relativistic heavy-ion
collisions.

\end{abstract}


\maketitle

\section{Introduction}
Relativistic fluid dynamics finds applications in cosmology,
astrophysics, and the physics of high-energy heavy-ion collisions. In
cosmology and certain areas of astrophysics, one needs a fluid
dynamics formulation consistent with the General Theory of Relativity
\cite{Ibanez}. On the other hand, a formulation based on the Special
Theory of Relativity is quite adequate to treat the evolution of the
strongly interacting matter formed in high-energy heavy-ion collisions
when it is close to a local thermodynamic equilibrium. The correct
formulation of the relativistic dissipative fluid dynamics is far from
settled and is currently under intense investigation
\cite{Baier:2006um,Baier:2007ix,Bhattacharyya:2008jc,Natsuume:2007ty,
El:2008yy,El:2009vj,Denicol:2010xn,Denicol:2012cn}.

In applications of fluid dynamics it is natural to first employ the
zeroth order (in gradients of the hydrodynamic four-velocity, for
example) or ideal fluid dynamics. However, as all fluids are dissipative
in nature due to the uncertainty principle
\cite{Danielewicz:1984ww}, the ideal fluid results serve only as a
benchmark when dissipative effects become important. The first-order
dissipative fluid dynamics or the relativistic Navier-Stokes (NS)
theory \cite{Landau} involves parabolic differential equations and
suffers from acausality and instability. The second-order
Israel-Stewart (IS) theory \cite{Israel:1979wp}, with its hyperbolic
equations restores causality but may not guarantee stability
\cite{Huovinen:2008te}.

The second-order viscous hydrodynamics has been quite successful in
explaining the spectra and azimuthal anisotropy of particles produced
in heavy-ion collisions at the Relativistic Heavy Ion Collider (RHIC)
\cite{Romatschke:2007mq,Song:2010mg} and recently at the Large Hadron
Collider (LHC) \cite{Luzum:2010ag,Qiu:2011hf}. However, IS theory can
lead to unphysical effects such as reheating of the expanding medium
\cite{Muronga:2003ta} and to a negative pressure
\cite{Martinez:2009mf} at large viscosity indicating its
breakdown. Furthermore, from comparison to the transport theory it was
demonstrated \cite{Huovinen:2008te,El:2008yy} that IS approach becomes
marginal when the shear viscosity to entropy density ratio $\eta/s
\gtrsim 1.5/(4\pi)$. With this motivation, the dissipative
hydrodynamic equations were extended \cite{El:2009vj} to third order,
which led to an improved agreement with the kinetic theory even for
moderately large values of $\eta/s$.

It is well known that the approach based on the generalized second law
of thermodynamics fails to capture all the terms in the evolution
equations of the dissipative quantities when compared with similar
equations derived from transport theory \cite{Baier:2006um}. It was
pointed out that using directly the definitions of the dissipative
currents, instead of the second moment of the Boltzmann equation as in
IS theory, one obtains identical equations of motion but with
different coefficients \cite{Denicol:2010xn}. Recently, it has been
shown \cite{Denicol:2012cn} that a generalization of Grad's 14-moment
method \cite{Grad} results in additional terms in the dissipative
equations.

It is important to note that all formulations that employ the
Boltzmann equation make a strict assumption of a local collision term
in the configuration space \cite{Israel:1979wp,Denicol:2010xn}. In
other words, within an infinitesimal fluid element containing a large
number of particles and extending over many interparticle spacings
\cite{Landau}, the different collisions that increase or decrease the
number of particles with a given momentum $p$ are all assumed to occur
at the same point $x^\mu$. This makes the collision integral a purely
local functional of the single-particle phase-space distribution
function $f(x,p)$ independent of the derivatives $\partial_\mu
f(x,p)$. In kinetic theory, $f(x,p)$ is assumed to vary slowly over
space-time, i.e., it changes negligibly over the range of
interparticle interaction \cite{deGroot}. However, its variation over
the fluid element may not be insignificant; see
Fig. \ref{flel}. Inclusion of the gradients of $f(x,p)$ in the
collision term will affect the evolution of dissipative quantities and
thus the entire dynamics of the system.

In this Letter, we shall provide a new formal derivation of the
dissipative hydrodynamic equations within kinetic theory but using a
nonlocal collision term in the Boltzmann equation. We obtain new
second-order terms and show that the coefficients of the other terms
are altered. These modifications do have a rather strong influence on
the evolution of the viscous medium as we shall demonstrate in the
case of one-dimensional scaling expansion.

\section{Nonlocal collision term}
Our starting point is the relativistic Boltzmann equation for the
evolution of the phase-space distribution function,
$p^\mu \partial_\mu f =  C[f]$,
where the collision term $C[f]$ is required to be consistent with the
energy-momentum and current conservation.  Traditionally $C[f]$ is
also assumed to be a purely local functional of $f(x,p)$, independent
of $\partial_\mu f$. This locality assumption is a powerful
restriction \cite{Israel:1979wp} which we relax by
including the gradients of $f(x,p)$ in $C[f]$. This
necessarily leads to the modified Boltzmann equation
\begin{equation}\label{MBE}
p^\mu \partial_\mu f = C_m[f] 
=  C[f] + \partial_\mu(A^\mu f) + \partial_\mu\partial_\nu(B^{\mu\nu}f) + \cdots,
\end{equation}
where $A^\mu$ and $B^{\mu\nu}$ depend on the type of the
collisions ($2 \leftrightarrow 2,~ 2 \leftrightarrow 3, \ldots$).

\begin{figure}[t]\begin{center}
\scalebox{0.35}{\includegraphics{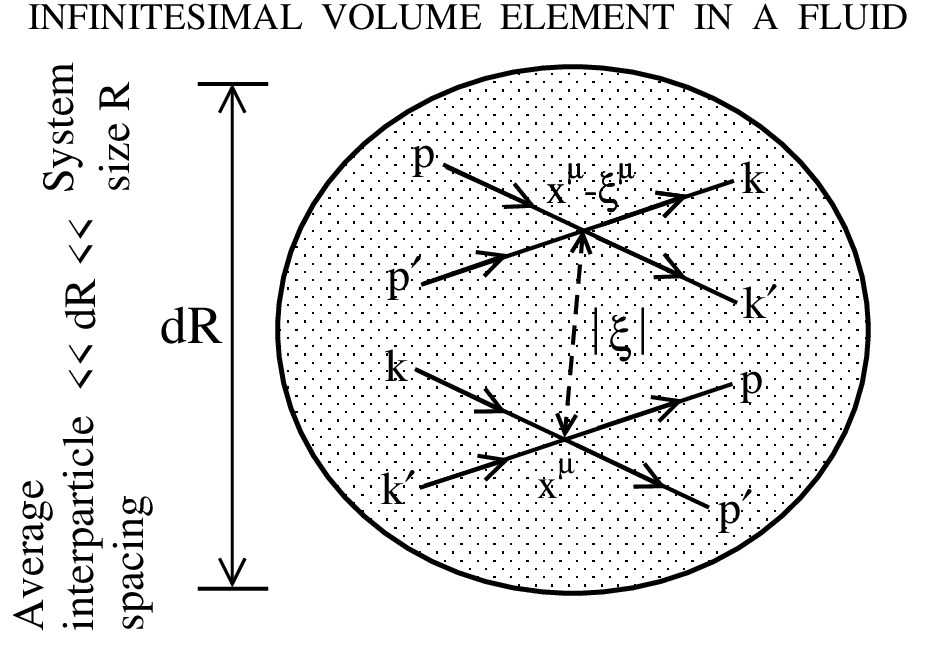}}
\end{center}
\vspace{-0.4cm}
\caption{Collisions $kk' \to pp'$ and $pp' \to kk'$ occurring at
  points $x^\mu$ and $x^\mu-\xi^\mu$ within an infinitesimal fluid
  element of size $dR$, around $x^\mu$, containing a large number of particles
  represented by dots.}
\label{flel}\end{figure}

For instance, for $2 \leftrightarrow 2$ 
elastic collisions,
\begin{eqnarray}\label{coll}
C[f]&=& \frac{1}{2} \int dp' dk \ dk' \  W_{pp' \to kk'} \nonumber \\
&& \times (f_k f_{k'} \tilde f_p \tilde f_{p'}
- f_p f_{p'} \tilde f_k \tilde f_{k'}),
\end{eqnarray}
where $ W_{pp' \to kk'}$ is the collisional transition rate, $f_p
\equiv f(x,p)$ and $\tilde f_p \equiv 1-r f(x,p)$ with $r = 1,-1,0$
for Fermi, Bose, and Boltzmann gas, and $dp = g d{\bf p}/[(2 \pi)^3
  \sqrt{{\bf p}^2+m^2}]$, $g$ and $m$ being the degeneracy factor and
particle rest mass. The first and second terms in Eq. (\ref{coll})
refer to the processes $kk' \to pp'$ and $pp' \to kk'$,
respectively. These processes are traditionally assumed to occur at
the same space-time point $x^\mu$ with an underlying assumption that
$f(x,p)$ is constant not only over the range of interparticle
interaction but also over the entire infinitesimal fluid element of
size $dR$, which is large compared to the average interparticle separation
\cite{Landau}; see Fig. \ref{flel}. Equation (1) together with this
crucial assumption has been used to derive the standard second-order
dissipative hydrodynamic equations
\cite{Romatschke:2009im,Israel:1979wp,Denicol:2010xn}. We, however,
emphasize that the space-time points at which the above two kinds
of processes
occur should be separated by an interval $|\xi^\mu| \leq dR$ within
the volume $d^4R$. It may be noted that the large number of particles
within $d^4R$ collide among themselves with various separations
$\xi^\mu$. Further, $\xi^\mu$ is independent of the
arbitrary point $x^\mu$ at which the Boltzmann equation is considered,
and is a function of $(p',k,k')$.
Of course, the points $(x^\mu-\xi^\mu)$ must lie within the past
light-cone of the point $x^\mu$ (i.e., $\xi^2 > 0$ and $\xi^0>0$) to
ensure that the evolution of $f(x,p)$ in Eq. (\ref{MBE}) 
does not violate causality.
With this realistic viewpoint, the second term in
Eq. (\ref{coll}) involves $f(x-\xi,p)f(x-\xi,p')\tilde
f(x-\xi,k)\tilde f(x-\xi,k')$, which on Taylor expansion at $x^\mu$ up
to second order in $\xi^\mu$, results in the modified Boltzmann
equation (\ref{MBE}) with
 \begin{eqnarray}\label{coeff1}
A^\mu &=& \frac{1}{2} \int dp' dk \ dk' \ \xi^\mu W_{pp' \to kk'}
f_{p'} \tilde f_k \tilde f_{k'}, \nonumber \\
B^{\mu\nu} &=& -\frac{1}{4} \int dp' dk \ dk' \ \xi^\mu \xi^\nu W_{pp' \to kk'}
f_{p'} \tilde f_k \tilde f_{k'}.
\end{eqnarray}

In general, for all collision types ($2 \leftrightarrow 2,~ 2
\leftrightarrow 3, \ldots$), the momentum dependence of the
coefficients $A^\mu$ and $B^{\mu\nu}$ can be made explicit by
expressing them in terms of the available tensors $p^\mu$ and the
metric $g^{\mu\nu} \equiv {\rm diag}(1,-1,-1,-1)$ as $A^\mu = a(x)p^\mu$
and $B^{\mu\nu}= b_1(x) g^{\mu\nu} + b_2(x) p^\mu p^\nu$, in the
spirit of Grad's 14-moment approximation. Equation
(\ref{MBE}) forms the basis of our derivation of the second-order
dissipative hydrodynamics.

\section{Hydrodynamic equations}
The conserved particle current and the energy-momentum tensor are
expressed as \cite{deGroot}
\begin{equation}\label{NT}
N^\mu = \int dp \ p^\mu f, ~~ T^{\mu\nu} = \int dp \ p^\mu p^\nu f. 
\end{equation}
The standard tensor decomposition of the above quantities results in
\begin{eqnarray}\label{NTD}
N^\mu &=& nu^\mu + n^\mu,  \nonumber\\
T^{\mu\nu} &=& \epsilon u^\mu u^\nu-(P+\Pi)\Delta ^{\mu \nu} 
+ \pi^{\mu\nu},
\end{eqnarray}
where $P, n, \epsilon$ are respectively pressure, number density,
energy density, and $\Delta^{\mu\nu}=g^{\mu\nu}-u^\mu u^\nu$ is the
projection operator on the three-space orthogonal to the hydrodynamic
four-velocity $u^\mu$ defined in the Landau frame: $T^{\mu\nu}
u_\nu=\epsilon u^\mu$. For small departures from equilibrium, $f(x,p)$
can be written as $f = f_0 + \delta f$. The equilibrium distribution
function is defined as $f_0 = [\exp(\beta u\cdot p -\alpha) + r]^{-1}$
where the inverse temperature $\beta=1/T$ and $\alpha=\beta\mu$ ($\mu$
being the chemical potential) are defined by the equilibrium matching
conditions $n\equiv n_0$ and $\epsilon \equiv \epsilon_0$. The scalar 
product is defined as $u.p\equiv u_\mu p^\mu$. The
dissipative quantities, viz., the bulk viscous pressure, the particle
diffusion current and the shear stress tensor are
\begin{eqnarray}\label{DISS}
\Pi &=& -\frac{\Delta_{\alpha\beta}}{3}\int dp \ p^
\alpha p^\beta \delta f,  \nonumber\\
n^\mu &=&  \Delta^{\mu\nu} \int dp \ p_\nu \delta f, \nonumber\\
\pi^{\mu\nu} &=& \Delta^{\mu\nu}_{\alpha\beta} \int dp \ p^
\alpha p^\beta \delta f.
\end{eqnarray}
Here
$\Delta^{\mu\nu}_{\alpha\beta} = [\Delta^\mu_\alpha \Delta^\nu_\beta +
  \Delta^\mu_\beta \Delta^\nu_\alpha - (2/3)\Delta^{\mu\nu}
  \Delta_{\alpha\beta}]/2$ is the traceless symmetric projection
operator. Conservation of current, $\partial_\mu N^\mu=0$ and energy-momentum tensor, $\partial_\mu
T^{\mu\nu} =0$, yield the fundamental evolution equations for $n$,
$\epsilon$ and $u^\mu$
\begin{eqnarray}\label{evol}
Dn+n\partial_\mu u^\mu + \partial_\mu n^\mu &=& 0, \nonumber \\
D\epsilon + (\epsilon+P+\Pi)\partial_\mu u^\mu - \pi^{\mu\nu}\nabla_{(\mu} u_{\nu)} &=& 0,  \nonumber\\
(\epsilon+P+\Pi)D u^\alpha - \nabla^\alpha (P+\Pi) + \Delta^\alpha_\nu \partial_\mu \pi^{\mu\nu}  &=& 0.
\end{eqnarray}
We use the standard notation $A^{(\alpha}B^{\beta )} = (A^\alpha
B^\beta + A^\beta B^\alpha)/2$, $D=u^\mu\partial_\mu$, and
$\nabla^\alpha = \Delta^{\mu\alpha}\partial_\mu$.
For later use we introduce 
$X^{\langle \mu \rangle}=\Delta^\mu_\nu X^\nu$ and
$X^{\langle \mu \nu \rangle}
=\Delta^{\mu \nu}_{\alpha \beta} X^{\alpha \beta}$.

Conservation of current and energy-momentum 
implies vanishing zeroth and first moments of the
collision term $C_m[f]$ in Eq. (\ref{MBE}), i.e., $\int dp \ C_m[f] =
0 = \int dp \ p^\mu C_m[f]$. Moreover, the arbitrariness in $\xi^\mu$
requires that these conditions be satisfied at each order in
$\xi^\mu$. Retaining terms up to second order in derivatives
leads to three constraint equations for the
coefficients ($a, b_1, b_2$), namely $\partial_\mu a = 0$,
\begin{eqnarray}\label{param}
\partial^2\left(b_1 \langle1\rangle_0 \right) 
+ \partial_\mu \partial_\nu\left( b_2 \langle p^\mu p^\nu\rangle_0\right) &=& 0, \nonumber\\
u_\alpha \partial_\mu \partial_\nu \left( b_2 \langle p^\mu p^\nu p^\alpha\rangle_0 \right)  
+ u_\alpha \partial^2 \left(b_1 n u^\alpha \right) &=& 0, 
\end{eqnarray}
where we define $\langle\cdots\rangle_0= \int dp (\cdots) f_0$. It is straightforward to show using
Eq. (\ref{param}) that the validity of the second law of thermodynamics, $\partial_\mu
s^\mu \geq 0$, enforces a further constraint $|a| < 1$,
on the collision term $C_m[f]$.

In order to obtain the evolution equations for the dissipative
quantities, we follow the approach as described by Denicol-Koide-Rischke
(DKR) 
in  Ref. 
\cite{Denicol:2010xn}. 
This approach employs directly the definitions of the dissipative
currents in contrast to the IS derivation which uses the second moment
of the Boltzmann equation.
The comoving derivatives of the dissipative 
quantities can be written from their definitions, 
Eq. (\ref{DISS}), as
\begin{eqnarray}\label{BE2}
\dot\Pi &=& -\frac{\Delta_{\alpha\beta}}{3}\int dp \ 
p^\alpha p^\beta \delta\dot f,  \nonumber\\
\dot n^\mu &=&  \Delta^{\mu\nu} \int dp \ p_\nu \delta\dot f, \nonumber\\
\dot\pi^{\mu\nu} &=& \Delta^{\mu\nu}_{\alpha\beta} \int dp \ 
p^\alpha p^\beta \delta\dot f,
\end{eqnarray}
where, $\dot X = D X$. Comoving derivative of the nonequilibrium part of 
the distribution function, $\delta\dot f$, can be obtained by writing the 
Boltzmann equation (\ref{MBE}) in the form,
\begin{equation}
\delta\dot f = - \dot f_0 - \frac{1}{u.p}p^\mu\nabla_\mu f + \frac{1}{u.p}C_m[f].
\end{equation}

To proceed further, we take recourse to Grad's 14-moment approximation
\cite{Grad} for the single-particle distribution in orthogonal basis
\cite{Denicol:2010xn}
\begin{eqnarray}\label{G14}
f = f_0 + f_0 \tilde f_0 \left( \lambda_\Pi \Pi + \lambda_n n_\alpha p^\alpha 
+ \lambda_\pi \pi_{\alpha\beta} p^\alpha p^\beta \right).
\end{eqnarray}
The coefficients ($\lambda_\Pi, \lambda_n, \lambda_\pi$) are functions
of ($n,\epsilon,\beta,\alpha$). Using Eqs. (\ref{BE2})-(\ref{G14}) and 
introducing first-order shear tensor
$\sigma_{\mu\nu}=\nabla_{\langle \mu}u_{\nu \rangle}$, vorticity
$\omega_{\mu \nu}=(\nabla_\mu u_\nu-\nabla_\nu u_\mu)/2$ and expansion
scalar $\theta=\partial \cdot u$, we finally obtain the following
evolution equations for the dissipative fluxes defined in
Eq. (\ref{DISS}):
\begin{align}
\dot\Pi = &~ -\frac{\Pi}{\tau_\Pi'}
- \beta_\Pi' \theta
- \tau_{\Pi n}' n \cdot \dot u 
- l_{\Pi n}' \partial \cdot n
- \delta_{\Pi\Pi}' \Pi\theta  \nonumber\\
&- \lambda_{\Pi n}' n \cdot \nabla \alpha
+ \lambda_{\Pi\pi}' \pi_{\mu\nu} \sigma^{\mu\nu}
+ \Lambda_{\Pi\dot u} \dot u \cdot \dot u \nonumber\\
& + \Lambda_{\Pi\omega} \omega_{\mu\nu} \omega^{\nu\mu} + (8 \ {\rm terms}) , \label{bulk}\\
\dot n^{\langle\mu\rangle} = &~ -\frac{n^\mu}{\tau_n'}
+ \beta_n' \nabla^\mu\alpha
- \lambda_{n\omega}' n_\nu \omega^{\nu\mu}
- \delta_{nn}' n^\mu \theta \nonumber \\
&- l_{n \Pi}'\nabla^\mu \Pi
+ l_{n \pi}'\Delta^{\mu\nu} \partial_\gamma \pi^\gamma_\nu
+ \tau_{n \Pi}' \Pi \dot u^\mu \nonumber \\
&- \tau_{n \pi}'\pi^{\mu \nu} \dot u_\nu
-\lambda_{n\pi}'n_\nu \pi^{\mu \nu}
+ \lambda_{n \Pi}'\Pi n^\mu \nonumber \\
&+  \Lambda_{n \dot u} \omega^{\mu \nu} \dot u_\nu
+ \Lambda_{n \omega} \Delta^\mu_\nu \partial_\gamma \omega^{\gamma \nu}
+ (9 \ {\rm terms}), \label{heat}\\
\dot\pi^{\langle\mu\nu\rangle} =&~ -\frac{\pi^{\mu\nu}}{\tau_\pi'}
+2 \beta_\pi' \sigma^{\mu\nu}
- \tau_{\pi n}' n^{\langle\mu}\dot u^{\nu\rangle}
+ l_{\pi n}' \nabla^{\langle \mu}n^{\nu\rangle} \nonumber\\
&+ 2\lambda_{\pi\pi}' \pi_\rho^{\langle \mu} \omega ^{\nu\rangle \rho}
+ \lambda_{\pi n}' n^{\langle\mu} \nabla^{\nu\rangle} \alpha
- \tau_{\pi\pi}' \pi_\rho^{\langle\mu} \sigma^{\nu\rangle\rho} \nonumber\\
&- \delta_{\pi\pi}' \pi^{\mu\nu}\theta
+ \Lambda_{\pi\dot u} \dot u^{\langle \mu} \dot u^{\nu\rangle}
+ \Lambda_{\pi\omega} \omega_\rho^{\langle \mu} \omega^{\nu\rangle\rho} \nonumber\\
&+ \chi_1 \dot b_2 \pi^{\mu\nu}
+ \chi_2 \dot u^{\langle \mu} \nabla^{\nu\rangle} b_2
+ \chi_3 \nabla^{\langle \mu} \nabla^{\nu\rangle} b_2. \label{shear}
\end{align}
The ``8 terms" (``9 terms'') involve second-order, linear
scalar (vector) combinations of derivatives of $b_1,b_2$.
All the terms in the above equations are inequivalent,
i.e., none can be expressed as a combination of others via equations
of motion \cite{Bhattacharyya:2012ex}. All the coefficients 
in Eqs. (\ref{bulk})-(\ref{shear})
are obtained as functions of hydrodynamic variables. For example,
some of the transport coefficients related to shear are
\begin{align}\label{coeff}
\tau_\pi' = \; & \beta_{\dot\pi}\tau_\pi,
~~~ \beta_\pi' = \tilde a \beta_\pi/\beta_{\dot\pi},
\nonumber\\
\beta_{\dot\pi} = \; & \tilde a + \frac{b_2}{3\eta\tilde a}
\left[\langle(u.p)^3\rangle_0-m^2n\right], 
\nonumber\\  
\beta_\pi = \; & \frac{4}{5}P + \frac{1}{15}(\epsilon-3P) - \frac{m^4}{15}\left<(u.p)^{-2}\right>_0,
\end{align}
where ${\tilde a}=(1-a)$. The rest of
the coefficients will be given in \cite{Amaresh_paper2}.

Retaining only the first-order terms in
Eqs. (\ref{bulk})-(\ref{shear}), and using DKR values of bulk 
viscosity $\zeta$, thermal conductivity $\kappa$ and shear
viscosity $\eta$, we get the modified first-order
equations for bulk pressure $\Pi = -\tau_{\Pi}' \beta_{\Pi}' \theta 
= - \tilde a \zeta \theta$, heat current $n^\mu = 
\beta_n' \tau_n' \nabla^\mu \alpha $ and shear stress tensor
$\pi^{\mu\nu} = 2 \tau_\pi'
\beta_\pi' \sigma^{\mu\nu} = 2 \tilde a \tau_\pi
\beta_\pi \sigma^{\mu\nu} = 2 \eta {\tilde a} \sigma^{\mu\nu}$. 
Thus the nonlocal collision term modifies even the first-order dissipative
equations. This constitutes one of the main results in the present
study. 

If $a,\ b_1$ and $b_2$ are all set to zero,
Eqs. (\ref{bulk})-(\ref{shear}) reduce to those obtained by DKR
\cite{Denicol:2010xn} with the same coefficients. Otherwise
coefficients of all the terms occurring in the DKR equations get
modified. Furthermore, our derivation results in new terms, for
instance those with coefficients $\Lambda_{k \dot u}$, $\Lambda_{k
  \omega}$, ($k=\Pi,n,\pi$), which are absent in \cite{Denicol:2010xn}
as well as in the standard Israel-Stewart approach
\cite{Israel:1979wp}. Hence these terms have also been missed so far
in the numerical studies of heavy-ion collisions in the hydrodynamic framework
\cite{Romatschke:2007mq,Song:2007ux,Luzum:2010ag}. Indeed
Eqs. (\ref{bulk})-(\ref{shear}) contain all possible second-order
terms allowed by symmetry considerations
\cite{Bhattacharyya:2012ex}. This is a consequence of the nonlocality
of the collision term $C_m[f]$. However, we note that a generalization
of the 14-moment approximation is also able to generate all these
terms as shown recently in Ref. \cite{Denicol:2012cn}.

\begin{figure}[t]\begin{center}
\scalebox{0.35}{\includegraphics{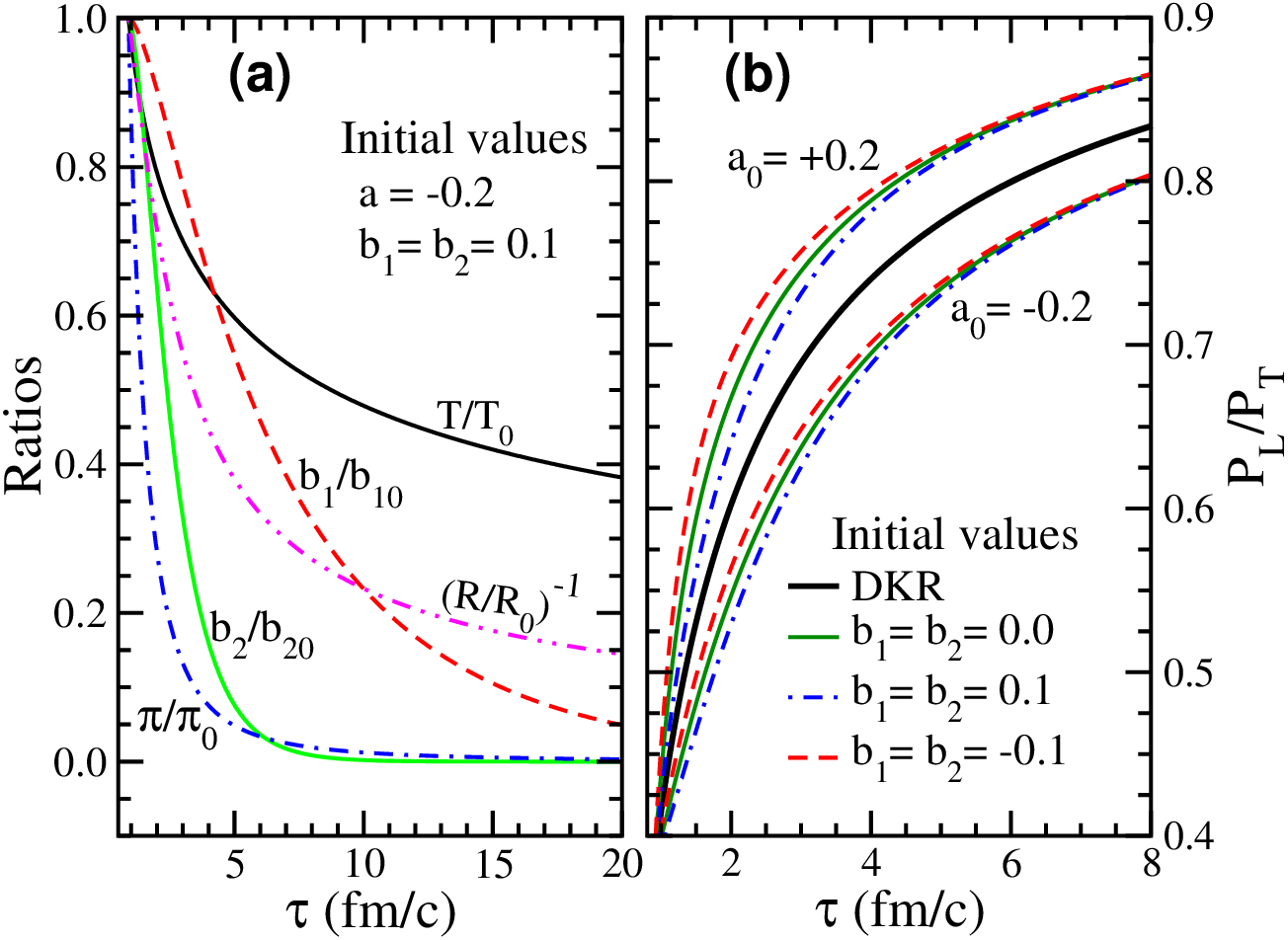}}
\end{center}
\vspace{-0.4cm}
\caption{Time evolution of (a) temperature, shear pressure, inverse
  Reynolds number and parameters ($b_1, \ b_2$) normalized to their
  initial values, and (b) anisotropy parameter $P_L/P_T$. Initial
  values are $\tau_0= 0.9$ fm/c, $T_0=360$ MeV, $\eta/s = 0.16$,
  $\pi_0 = 4\eta/(3\tau_0)$. Units of $b_2$ are GeV$^{-2}$. The curve
  labelled DKR is obtained by setting $a=b_1=b_2=0$ in
  Eqs. (\ref{Bjpi}) and (\ref{coeffu}).}
\label{evol1}\end{figure}

\section{Numerical results}
To demonstrate the numerical significance of the new dissipative
equations derived here, we consider evolution of a massless Boltzmann
gas, with equation of state $\epsilon=3P$, at vanishing net baryon
number density in the Bjorken model \cite{Bjorken:1982qr}. The new
terms, namely $\dot u \cdot \dot u$, $\omega_{\mu\nu}
\omega^{\nu\mu}$, $\omega^{\mu \nu} \dot u_\nu$, $\Delta^\mu_\nu
\partial_\gamma \omega^{\gamma \nu}$, $\dot u^{\langle \mu} \dot
u^{\nu\rangle}$ and $\omega_\rho^{\langle \mu}
\omega^{\nu\rangle\rho}$ containing acceleration and vorticity do not
contribute in this case. However, they are expected to play an
important role in the full 3D viscous hydrodynamics.

In terms of the coordinates ($\tau,x,y,\eta$) where $\tau =
\sqrt{t^2-z^2}$ and $\eta=\tanh^{-1}(z/t)$, the initial four-velocity
becomes $u^\mu=(1,0,0,0)$. In this scenario $\Pi=0=n^\mu$ and the
equation for $\pi \equiv -\tau^2 \pi^{\eta \eta}$ reduces to
\begin{eqnarray}\label{Bjpi}
\frac{\pi}{\tau_\pi} + \beta_{\dot\pi}\frac{d\pi}{d\tau} = 
\beta_\pi \frac{4}{3\tau} - \lambda \frac{\pi}{\tau} 
- \psi \pi \frac{db_2}{d\tau},
\end{eqnarray}
where the coefficients are
\begin{eqnarray}\label{coeffu}
\beta_{\dot\pi} \! &=& \! \tilde a + \frac{b_2 (\epsilon + P)}{\tilde a \beta\eta},
~~ \beta_\pi = \frac{4}{5} \tilde a P ,
~~ \psi = \frac{9 (\epsilon + P)}{5 \tilde a \beta \eta},
 \nonumber \\
\lambda \! &=& \! \frac{38}{21}\tilde a 
- \left( \frac{b_1\beta}{5} - \frac{8b_2}{7\beta}\right)\frac{\epsilon + P}{\tilde a \eta}.
\end{eqnarray}
For comparison we quote the IS results \cite{Israel:1979wp}:
$\beta_\pi=2P/3,~ \lambda=2$.
The coupled differential equations (\ref{evol}), (\ref{param}) and
(\ref{Bjpi}) are solved simultaneously for a variety of initial
conditions: temperature $T=360$ or 500 MeV corresponding to typical
RHIC and LHC energies, and shear pressure $\pi=0$ or $\pi=\pi_{\rm
  NS}=4 \eta/(3 \tau_0)$ corresponding to isotropic and anisotropic
pressure configurations. Since the nonlocal effects embodied in the
Taylor expansion (\ref{MBE}) are not large, the initial $a,~b_1,~b_2$
are so constrained that the corrections to first-order and second-order 
terms remain small; recall
also the additional constraints $|a| < 1$ and Eq. (\ref{param}).

\begin{figure}[t]\begin{center}
\scalebox{0.35}{\includegraphics{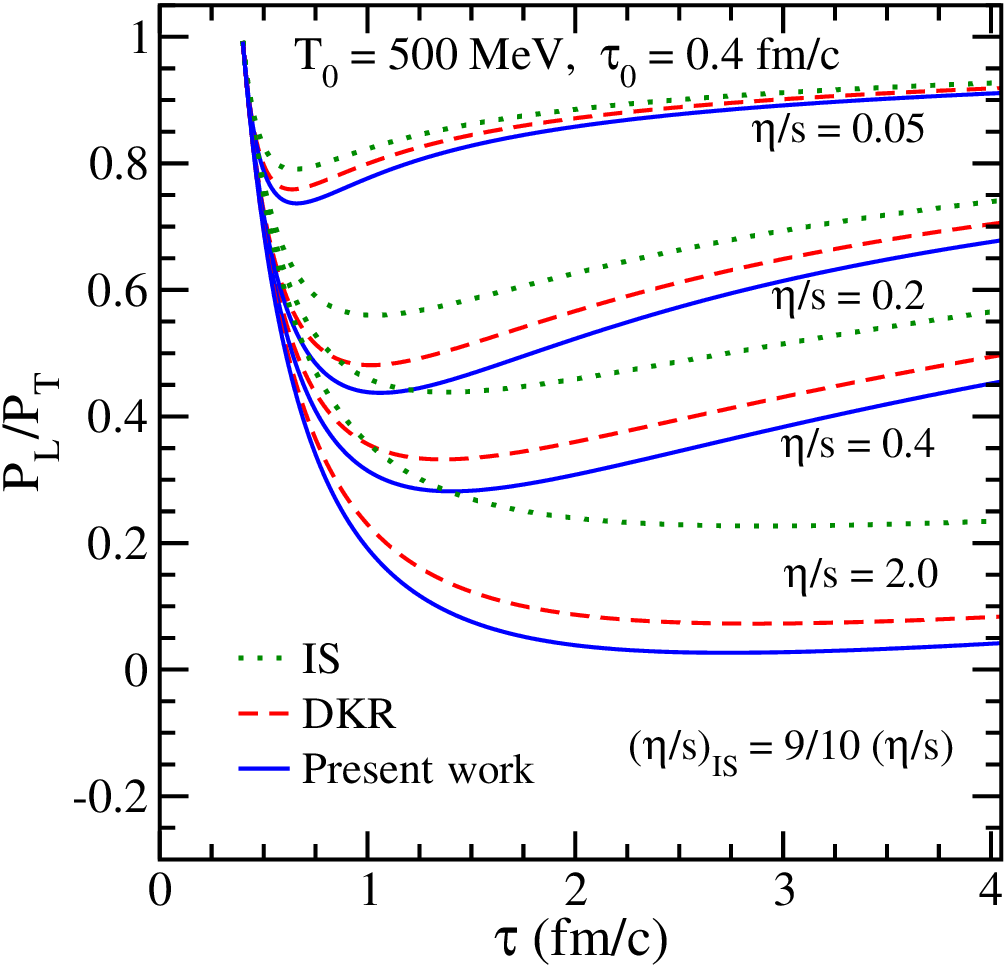}}
\end{center}
\vspace{-0.4cm}
\caption{Time evolution of $P_L/P_T$ in IS \cite{Israel:1979wp}, DKR ($a=b_1=b_2=0$), 
and the present work, for isotropic initial
  pressure configuration $(\pi_0=0)$. The scaling $(\eta/s)_{\rm IS}=9/10(\eta/s)$ 
ensures that all the results are compared at the same cross section 
\cite{Denicol:2010xn}.}
\label{evol2}\end{figure}

Figure 2(a) illustrates the evolution of these quantities for a choice of
initial conditions. $T$ decreases monotonically to the crossover
temperature $~170$ MeV at time $\tau \simeq 10$ fm/c, which is
consistent with the expected lifetime of quark-gluon plasma.
Parameter $a$ is constant whereas $b_1$ and $b_2$ vary smoothly and
tend to zero at large times indicating reduced but still significant
presence of nonlocal effects in the collision term at late times. This
is also evident in Fig. 2(b) where the pressure anisotropy
$P_L/P_T=(P-\pi)/(P+\pi/2)$ shows marked deviation from IS,
controlled mainly by $a$. At late times $P_L/P_T$ is largely
unaffected by the choice of initial values of $b_1,~b_2$. Although
the shear pressure $\pi$ vanishes rapidly indicating approach to ideal
fluid dynamics, the $P_L/P_T$ is far from unity. Faster isotropization
for initial $a>0$ may be attributed to a smaller effective shear
viscosity $(1-a) \eta$ in the modified NS equation, and
conversely. Figure 2(b) also indicates the convergence of the Taylor
expansion that led to Eq. (\ref{MBE}).

Figure 3 shows the evolution of $P_L/P_T$ for isotropic initial
pressure configuration, at various $\eta/s$ for the LHC energy regime.
Compared to IS, DKR leads to larger pressure anisotropy. Further, with
small initial corrections ($ 10$\% to first-order and $\simeq 20$\% to
the second-order terms) due to $a,\ b_1,\ b_2$, the nonlocal
hydrodynamics (solid lines) exhibits appreciable deviation from the
(local) DKR theory. The above results clearly demonstrate the
importance of the nonlocal effects, which should be incorporated in
transport calculations as well. Comparison of nonlocal hydrodynamics
to nonlocal transport theory would be illuminating.

In a realistic 2+1 or 3+1 D calculation, one has to choose the
thermalization time and the freeze-out temperature together with
suitable initial conditions for hydrodynamic velocity, energy density,
shear pressure as well as for the nonlocal coefficients $a, ~b_1,~
b_2$ to fit $dN/d\eta$ and $p_T$ spectra of hadrons, and then predict,
for example, the anisotropic flow $v_n$ for a given $\eta/s$. Nonlocal
effects (especially via $a$) will affect the extraction of $\eta/s$
from fits to the measured $v_n$. It may also be noted that although
(local) viscous hydrodynamics explains the gross features of $\pi^-$
and $K^-$ spectra for the (0-5)\% most central Pb-Pb collisions at
$\sqrt{s_{NN}}=2.76$ TeV, it strongly disagrees with the measured
$\bar p$ spectrum \cite{Floris:2011ru}. Further the constituent quark
number scaling violation has been observed in the $v_2$ and $v_3$ data
for $\bar p$, at this LHC energy \cite{Krzewicki:2011ee}. The above
discrepancies may be attributed partly to the nonlocal effects which
can have different implications for two- and three-particle
correlations and thus affect the meson and baryon spectra differently.

\section{Summary}
To summarize, we have presented a new derivation of the relativistic
dissipative hydrodynamic equations by introducing a nonlocal
generalization of the collision term in the Boltzmann equation. The
first-order and second-order equations are modified: new terms occur
and coefficients of others are altered. While it is well known that
the derivation based on the generalized second law of thermodynamics
misses some terms in the second-order equations, we have shown that
the standard derivation based on kinetic theory and 14-moment
approximation also misses other terms. The method presented here is
able to generate all possible terms to a given order that are allowed
by symmetry. It can also be extended to derive third-order
hydrodynamic equations.

\begin{acknowledgments}
We thank S. Bhattacharyya, J.-P. Blaizot, M. Luzum, S. Majumdar,
S. Minwalla and J.-Y. Ollitrault for helpful discussions. AJ thanks
G.S. Denicol and A. El for several useful correspondences.
\end{acknowledgments}

\end{document}